\documentclass[sigconf,screen]{acmart}
\usepackage{algorithm}
\usepackage{algorithmic}
\usepackage{color}
\usepackage{multirow}
\usepackage{pifont}
\usepackage{makecell}

\AtBeginDocument{%
  \providecommand\BibTeX{{%
    \normalfont B\kern-0.5em{\scshape i\kern-0.25em b}\kern-0.8em\TeX}}}

\begin{document}

\title{Mutation-Based Deep Learning Framework Testing Method in JavaScript Environment}

\author{Yinglong Zou}
\affiliation{%
  \institution{State Key Laboratory for Novel Software Technology\\ Nanjing University}
  \country{China}
}
\email{652023320004@smail.nju.edu.cn}
\orcid{0009-0006-9375-7417}

\author{Juan Zhai}
\affiliation{%
  \institution{University of Massachusetts Amherst 
  \country{United States}}
}
\email{juanzhai@umass.edu}
\orcid{0000-0001-5017-8016}

\author{Chunrong Fang}
\authornote{Chunrong Fang and Zhenyu Chen are the corresponding authors.}
\affiliation{%
  \institution{State Key Laboratory for Novel Software Technology\\ Nanjing University}
  \country{China}
}
\email{fangchunrong@nju.edu.cn}
\orcid{0000-0002-9930-7111}

\author{Jiawei Liu}
\affiliation{%
  \institution{State Key Laboratory for Novel Software Technology\\ Nanjing University}
  \country{China}
}
\email{jw.liu@smail.nju.edu.cn}
\orcid{0000-0002-4930-9637}

\author{Tao Zheng}
\affiliation{%
  \institution{State Key Laboratory for Novel Software Technology\\ Nanjing University}
  \country{China}
}
\email{zt@nju.edu.cn}
\orcid{0009-0001-3736-4604}

\author{Zhenyu Chen}
\authornotemark[1]
\affiliation{%
  \institution{State Key Laboratory for Novel Software Technology\\ Nanjing University}
  \country{China}
}
\email{zychen@nju.edu.cn}
\orcid{0000-0002-9592-7022}

\renewcommand{\shortauthors}{Yinglong Zou et al.}

\begin{abstract}
In recent years, Deep Learning (DL) applications in JavaScript environment have become increasingly popular. 
As the infrastructure for DL applications, JavaScript DL frameworks play a crucial role in the development and deployment. It is essential to ensure the quality of JavaScript DL frameworks. 
However, the bottleneck of limited computational resources in the JavaScript environment brings new challenges to framework testing. Specifically, JavaScript DL frameworks are equipped with various optimization mechanisms (e.g., cache reuse, inference acceleration) to overcome the bottleneck of limited computational resources. These optimization mechanisms are overlooked by existing methods, resulting in many bugs in JavaScript DL frameworks being missed. To address the above challenges, we propose a mutation-based JavaScript DL framework testing method named DLJSFuzzer.
DLJSFuzzer designs 13 tensor mutation rules targeting the cache reuse mechanism to generate test input tensors.
Besides, DLJSFuzzer designs eight model mutation rules targeting the inference acceleration mechanism to generate test input models.
To evaluate the effectiveness of DLJSFuzzer, we conduct experiments on the most widely-used JavaScript DL framework, TensorFlow.js. 
The experimental results show that DLJSFuzzer outperforms state-of-the-art methods in both effectiveness and efficiency. 
DLJSFuzzer successfully detects 21 unique crashes and 126 unique NaN \& Inconsistency bugs. 
All detected crashes have been reported to the open-source community, with 12 of them already confirmed by developers. 
Additionally, DLJSFuzzer has improved by over 47\% in model generation efficiency and over 91\% in bug detection efficiency compared to all baselines.
\end{abstract}


\maketitle

\section{Introduction}\label{introduction}

With the popularization of web applications, JavaScript has become the primary language for frontend development \cite{flanagan2011javascript}. As developers increasingly choose to integrate deep learning (DL) models into web applications, the demand for developing and deploying DL models in JavaScript environments is growing \cite{bileschi2020deep}.
The development and deployment of JavaScript DL applications rely on JavaScript DL frameworks (e.g., TensorFlow.js~\cite{tfjs}), which provide developers with various functional operators. Unlike common DL frameworks like TensorFlow \cite{tensorflow}, JavaScript DL frameworks have limited computational resources \cite{cache_reuse}.
To overcome this bottleneck, JavaScript DL frameworks design a series of optimization mechanisms (e.g. cache reuse, inference acceleration), which brings new challenges to DL framework testing in JavaScript environment. 
Unfortunately, existing methods failed to identify this important and unique challenge.
Once overlooked, the testing will miss certain bugs (e.g., crashes arising from inference acceleration, see Section \ref{RQ2_root cause}). 
The ignorance manifests in two main components of the existing methods: \textbf{test input model generation} and \textbf{test input tensor generation}.

In \textbf{test input tensor generation}, existing methods fail to flexibly adjust the shape and data types of test input tensors, making it challenging to effectively detect framework bugs that may arise from cache reuse mechanism. 
Specifically, existing DL framework testing methods define the shape and data types of test input tensors before testing and keep them unchanged throughout testing, which brings two problems. Firstly, the fixed shape and data types of the test input tensors limit the ability to thoroughly test the cache reuse mechanism (e.g., it fails to test modules that deal with varying tensor shapes). Secondly, the fixed shapes and data types of test input tensors can hardly trigger misjudgments in the framework, which overlooks related bugs. For example, JavaScript DL frameworks occasionally misjudge two tensors with identical shapes as completely identical tensors, leading to incorrect cache reuse (see Section \ref{background cache reuse} for more information about this misjudgment). 

In \textbf{test input model generation}, the models generated by existing methods lack model structures that can trigger inference acceleration, making it difficult to effectively detect framework bugs that may arise from inference acceleration mechanisms. Specifically, existing framework testing methods are designed for common DL frameworks like TensorFlow and PyTorch, without targeting the inference acceleration mechanism in the JavaScript environment. The models generated by existing methods hardly trigger inference acceleration strategies such as node optimization, operator reordering, and operator fusion. As a result, it is almost impossible to detect bugs in the JavaScript DL frameworks caused by inference acceleration by existing methods (see Section \ref{RQ2_root cause}).

To better test JavaScript DL framework, we propose a mutation-based method called DLJSFuzzer, which targets widely used optimization mechanisms (including cache reuse and inference acceleration) in JavaScript DL frameworks to generate test input tensors and models. 
DLJSFuzzer designs 13 tensor mutation rules to generate new test input tensors flexibly in the shape and data types, which contributes to testing the JavaScript DL framework components related to the cache reuse mechanism. 
In addition, DLJSFuzzer designs eight model mutation rules targeting three different inference acceleration strategies (including node optimization, operator reordering, and operator fusion) to generate new models successfully triggering the inference acceleration mechanism. 
We apply DLJSFuzzer to test the most widely used JavaScript DL framework, TensorFlow.js \cite{tfjs}, and compare it against three state-of-the-art methods (including LEMON \cite{lemon}, Muffin \cite{muffin}, and Gandalf \cite{gandalf}). Experimental results show that DLJSFuzzer outperforms all baselines in effectiveness, and achieves an over 47\% improvement in model generation efficiency and an over 91\% improvement in bug detection efficiency. 
DLJSFuzzer has successfully detected 21 unique crashes and 126 unique NaN \& Inconsistency bugs. All detected crashes have been reported to the open-source community, with 12 of them already confirmed by developers. The main root causes of the detected crashes include cache reuse, implementation bug, inference acceleration, and others. The main root causes of the detected NaN \& Inconsistency bugs include precision, environment, cache reuse, and randomness. However, all crashes detected by baselines stem from cache reuse, lacking those from implementation bug, inference acceleration, and others. All NaN \& Inconsistency bugs detected by baselines stem from random and precision, lacking those from environment and cache reuse.

Our main contributions are as follows:
\begin{itemize}
\item We propose a mutation-based JavaScript DL framework testing method, which is the first to target optimization mechanism in JavaScript DL frameworks.
\item We design 13 mutation rules for generating test input tensors targeting the cache reuse mechanism in the JavaScript environment, supporting the shape and data type of the test input tensors to change flexibly during method execution.
\item We design eight mutation rules for generating test input models targeting the inference acceleration mechanism in JavaScript DL frameworks, which contributes to generating more models covering inference acceleration strategies.
\item We implement the method as a tool named DLJSFuzzer and apply DLJSFuzzer to detect the most widely used JavaScript DL framework, TensorFlow.js. Experimental results show that DLJSFuzzer successfully detects 21 crashes and 126 NaN \& Inconsistency bugs. In addition, DLJSFuzzer achieves an over 47\% improvement in model generation efficiency and an over 91\% improvement in bug detection efficiency.
\end{itemize}

We make all the data and code used in our paper publicly available on Github: \url{https://github.com/DLJSFuzzer/DLJSFuzzer}.

\section{Background}
\subsection{DL in JavaScript Environment}
\label{DL in JavaScript Environment}
With the development of web browsers and the Node.js platform, the application of deep learning in the JavaScript environment is becoming increasingly popular. The JavaScript DL frameworks are the infrastructure for DL applications, which are designed to meet the need for developing and deploying DL models in the JavaScript environment. Among them, TensorFlow.js is the most commonly used JavaScript DL framework \cite{tfjsbackground}. Through TensorFlow.js, developers can implement applications in multiple areas such as image classification \cite{image_classification}, object detection \cite{object_detection}, natural language processing \cite{nlp}, and recommendation systems \cite{recommendation_systems}.

Despite the widespread application of JavaScript DL frameworks, they also have some limitations. Specifically, compared to the local environment, the deployment environment of JavaScript DL frameworks has more scarce computational resources. On the one hand, the browser cache size is limited. The browser cache is mainly used to temporarily store model parameters (including input tensors). The cache size of commonly used browsers is usually less than 100 MB, which is much smaller than the total size of parameters in the model (e.g., the parameter size of VGG reaches 140 MB). The significant disparity between the browser cache and the total size of model parameters results in excessive cache read and write operations during the model inference process, severely reducing the efficiency of model inference. On the other hand, the model inference speed is limited. JavaScript DL frameworks are often deployed on mobile devices (e.g., mobile phones). Compared to deploying in a local environment, the floating point arithmetic speed in mobile devices is more limited, leading to a slower model inference speed.

\subsection{Optimization in JavaScript DL Framework}
\label{Optimization in JavaScript DL Framework}

As introduced in Section \ref{DL in JavaScript Environment}, the scarcity of computing resources (e.g., limited browser cache and model inference speed) is a major bottleneck of developing and deploying DL models in the JavaScript environment.
To overcome this bottleneck, a series of optimization mechanisms are designed in JavaScript DL frameworks. Among them, \textbf{cache reuse} and \textbf{inference acceleration} are the most popular ones. In this section, we will take TensorFlow.js as an example to explain these optimization mechanisms in detail.

\subsubsection{cache reuse \cite{cache_reuse_2}}
\label{background cache reuse}
Unlike traditional cache reuse mechanism, 
JavaScript-environment cache reuse mechanism is designed to adapt the small browser cache in the JavaScript environment, which relies on complex browser cache refresh strategies. If the tensors that need to be newly loaded are identical to the ones already loaded, the browser will no longer load the new tensors, but will directly use the ones that have been loaded. However, the JavaScript DL framework may make misjudgments when determining if two tensors are completely identical. If the newly loaded tensor has the same dimensions as the already loaded tensor, it may be mistakenly considered identical. This misjudgment can lead to loading parameters that have the same dimensions but are not completely identical, resulting in framework crashes or incorrect model inference results.

\subsubsection{inference acceleration \cite{inference_acceleration}}
\label{background inference acceleration}

TensorFlow.js primarily accelerates inference by optimizing the computation graph corresponding to the model. The computation graph is a directed acyclic graph that represents the operators in the model and the data flow corresponding to the model. Common inference acceleration strategies include \textbf{node optimization}, \textbf{operator reordering}, and \textbf{operator fusion}. A more detailed introduction to these inference acceleration strategies is as follows.

\begin{figure}[htpb]
    \centering
    \includegraphics[scale=0.05]{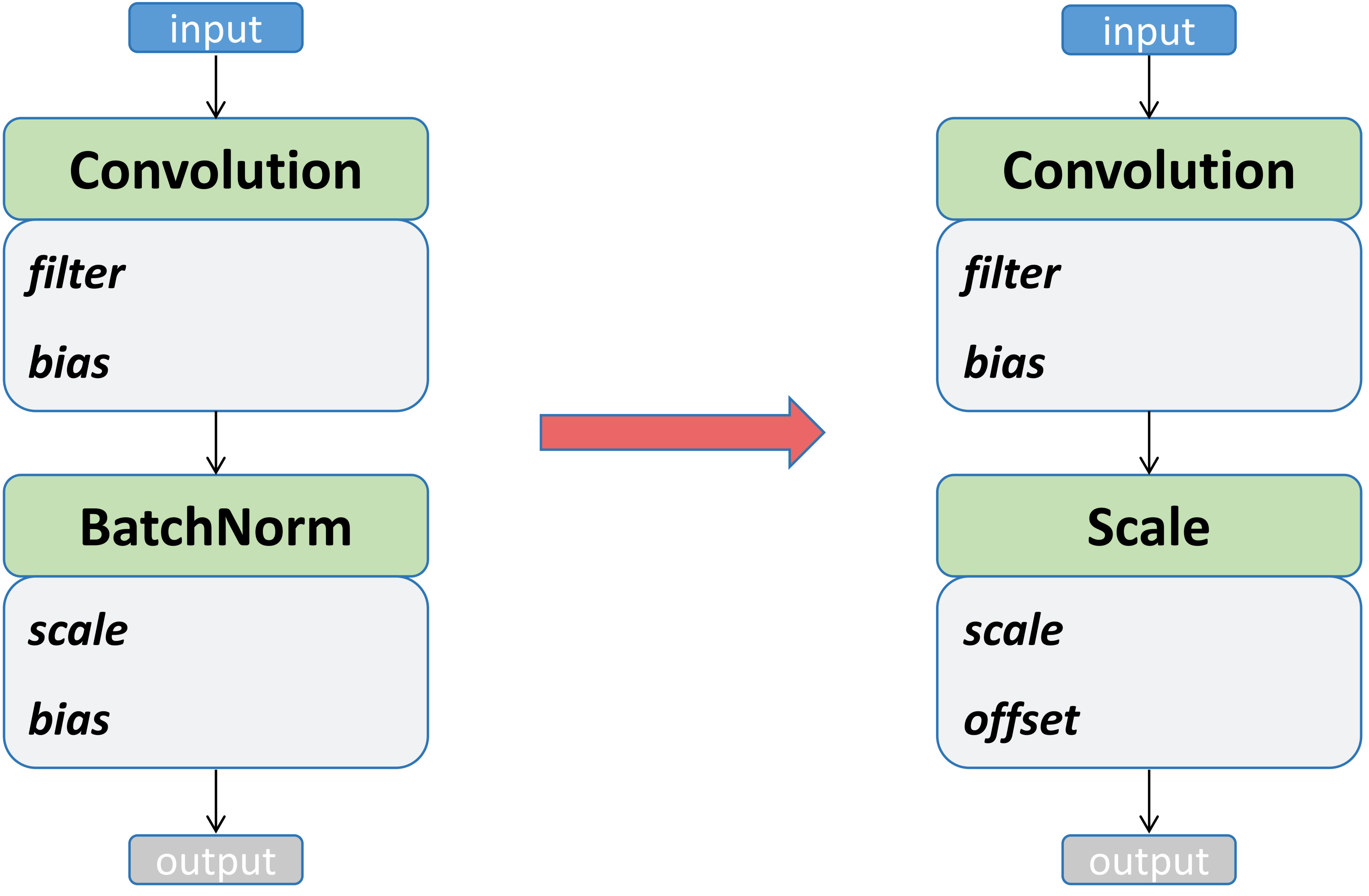}
    \caption{An Example of Node Optimization}
    \label{figure_node optimization}
\end{figure}

\textbf{Node Optimization.}
Node optimization aims to replace nodes in the model with nodes that have the same computational logic but are more suitable for deployment in a JavaScript environment. In the DL model, a node is a combination of an operator and tensors \& parameters corresponding to this operator. Node optimization simplifies the computational formulas corresponding to nodes and maps the simplified formulas to more efficient nodes. Figure \ref{figure_node optimization} shows an example of node optimization. The rectangles in this figure represent nodes, which include operators and their corresponding parameters. Node optimization replaces nodes corresponding to the operator $Batchnorm$ with nodes corresponding to the operator $Scale$. In this way, it is possible to reduce computational complexity, decrease the model's parameter size, and alleviate the insufficient browser cache in the JavaScript environment. 

\begin{figure}[htpb]
    \centering
    \includegraphics[scale=0.05]{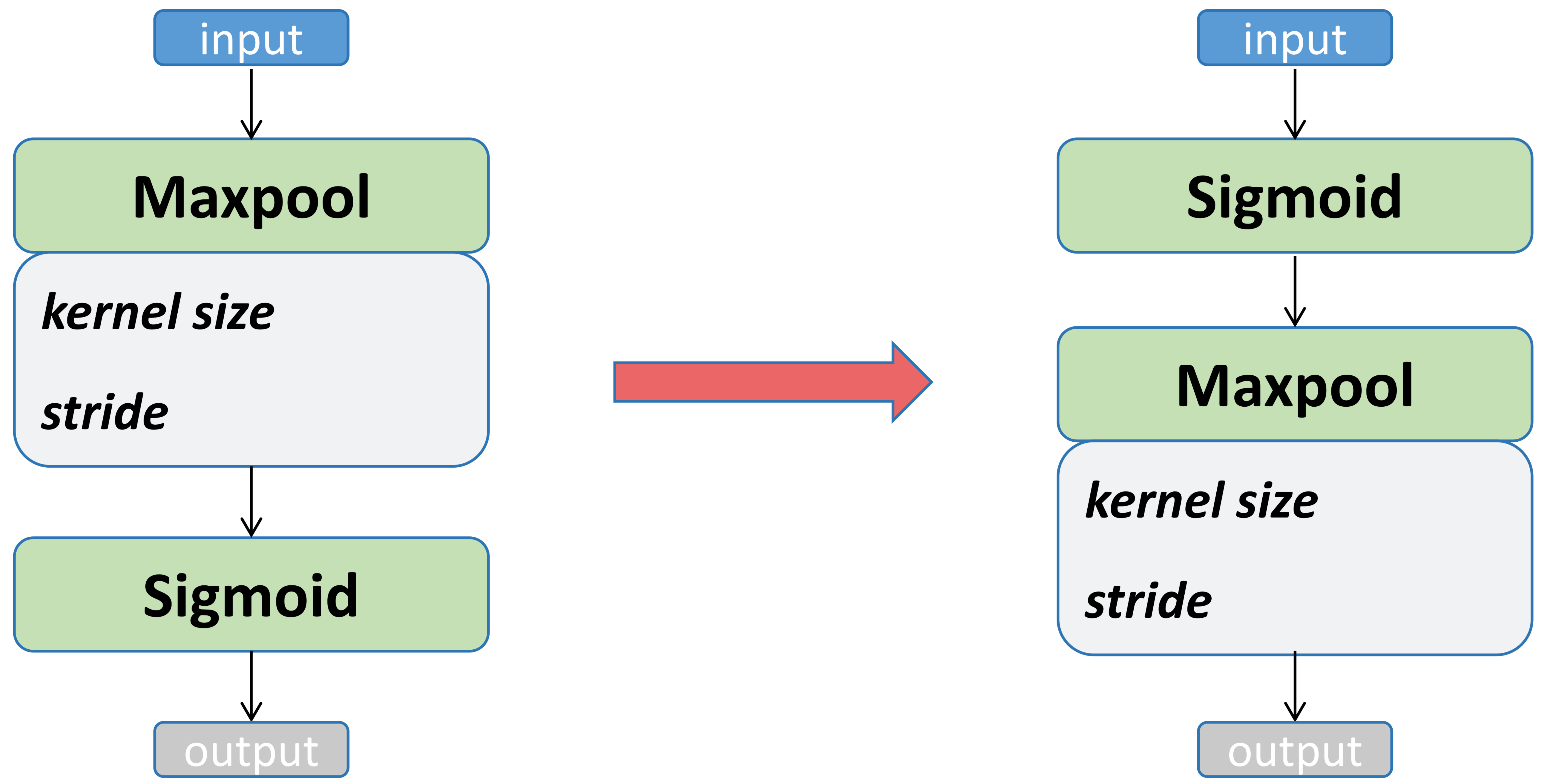}
    \caption{An Example of Operator Reordering}
    \label{figure_operator reordering}
\end{figure}

\textbf{Operator Reordering.} Operator reordering reduces the computational workload of model inference by rearranging the topological order of operators in the model while maintaining the same inference accuracy. The goal of operator reordering is to improve the efficiency of model inference by reorganizing the computational sequence of operators, reducing unnecessary computations and data transfers. Figure \ref{figure_operator reordering} shows an example of operator reordering. In this example, the computational process is simplified by exchanging the order of the operator $MaxPool$ and the operator $Sigmoid$.

\begin{figure}[htpb]
    \centering
    \includegraphics[scale=0.05]{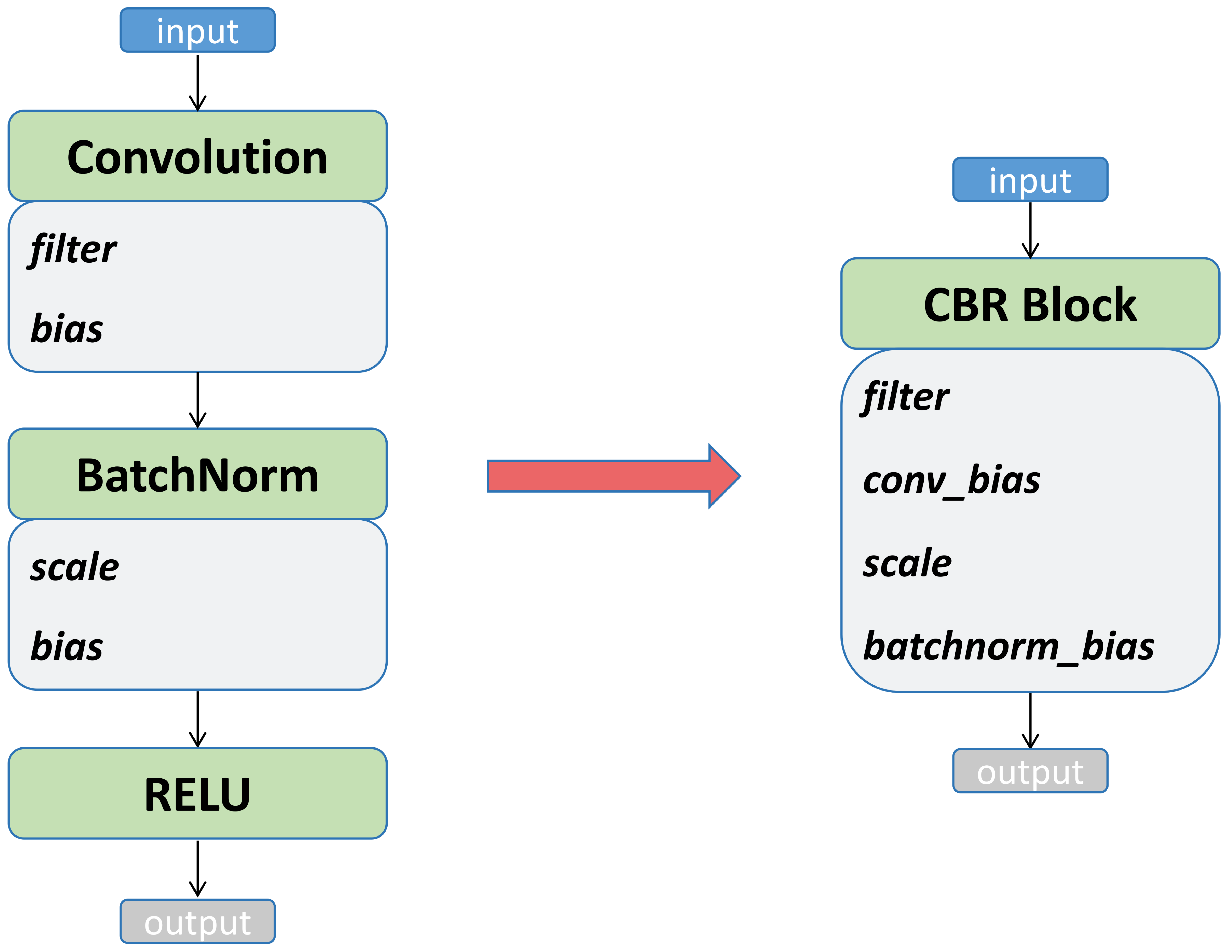}
    \caption{An Example of Operator Fusion}
    \label{figure_operator fusion}
\end{figure}

\textbf{Operator Fusion.}
Operator fusion is a model compression technique that merges multiple related operators into a single operator based on the equivalence relationship between operators and operator combinations. In DL models, a large number of repeated calculations and frequent memory accesses lead to inference latency and energy consumption issues. Operator fusion analyzes the model's topology, identifies operators that can be merged, and combines them into a new operator according to certain rules. Operator fusion can effectively reduce the number of nodes and edges in the computation graph corresponding to the model. Reducing the number of nodes means fewer memory accesses, and reducing the number of edges means fewer computational steps. Therefore, operator fusion can reduce the computational complexity and storage usage, thereby accelerating the model's inference. Figure \ref{figure_operator fusion} shows an example of operator fusion. In this example, the operator sequence $(Convolution, BatchNorm, RELU)$ is fused to a CBR (Conv\_Batch\_Relu) block.

Unlike traditional inference acceleration, JavaScript-environment inference acceleration employs new optimization strategies (e.g., In-place Node Optimization) and unique implementations of common strategies (e.g., depending on the browser API), introducing unique types of bugs to JavaScript-environment DL frameworks (see Section \ref{RQ2_root cause}).
\section{Methodology}
\subsection{Overview}
\label{overview}
\begin{figure*}[htpb]
    \centering
    \includegraphics[scale=0.075]{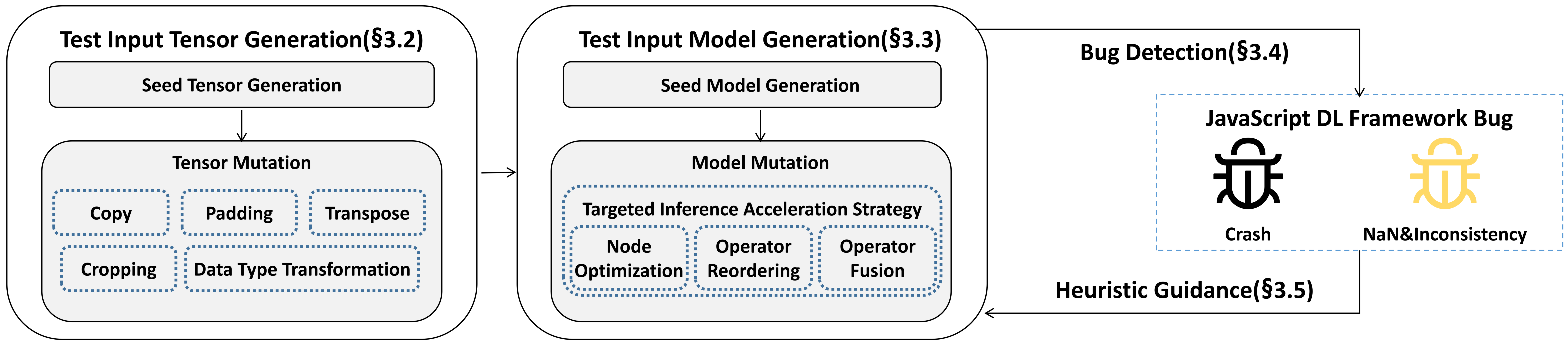}
    \Description{workflow}
    \caption{The Workflow of DLJSFuzzer}
    \label{workflow}
\end{figure*}

In this work, we propose DLJSFuzzer, a mutation-based JavaScript DL framework testing method. Figure \ref{workflow} shows the overall workflow of DLJSFuzzer. Firstly, DLJSFuzzer generates seed tensors and designs tensor mutation rules targeting the cache reuse mechanism to generate new test input tensors flexibly in the shape and data types (Section \ref{test input tensor generation}). Secondly, DLJSFuzzer generates a seed model and designs model mutation rules targeting inference acceleration strategies (e.g., node optimization, operator reordering, and operator fusion) to generate new test input models  (Section \ref{test input model generation}). Based on generated test inputs, DLJSFuzzer detects DL framework bugs by analyzing the testing results under different frameworks and browsers (Section \ref{bug detection}), and heuristically guides the next round based on these results (Section \ref{heuristic guidance}).

\subsection{Test Input Tensor Generation}
\label{test input tensor generation}
The test input tensor is an important part of the test input in DL framework testing. The cache reuse mechanism applied in JavaScript DL frameworks poses new challenges for generating test input tensors. Some false positives usually occur when judging whether two tensors are exactly the same. These false positives may lead to incorrect inference results or even cause crashes in DL frameworks (see Section \ref{Optimization in JavaScript DL Framework}). However, these quality issues cannot be successfully detected by existing DL framework testing methods. Specifically, existing DL framework testing methods require determining the shape and data type of the test input tensors before the testing method starts. Throughout the testing process, the shapes and data types of all test input tensors remain almost (if not completely) unchanged. The above process has the following two limitations. On the one hand, compared to the flexible and variable test input tensors in terms of shape and data types, the test input tensors generated by existing methods cannot fully and comprehensively test the cache reuse mechanism (e.g., modules related to handling different tensor shapes are not tested). On the other hand, test input tensors generated by existing methods with unchanged shapes and data types almost cannot trigger the cache reuse mechanism to misjudge whether two tensors are completely identical, thus overlooking relevant DL framework quality issues.

To overcome these limitations, DLJSFuzzer designs a mutation-based test input tensor generation method, which proposes a series of tensor mutation strategies to flexibly change the shape and data type of test input tensors. The detailed method is as follows: Firstly, DLJSFuzzer generates a seed tensor in the format of (N, C, H, W). DLJSFuzzer supports both randomly generating the seed tensor and loading the seed tensor from open-source datasets (e.g., MNIST). All adopted open-source datasets are listed in Section \ref{experimental Setup}. After that, DLJSFuzzer randomly selects a tensor mutation rule to change the shape and data types of the test input tensor. Table \ref{tensor mutation rule} shows all tensor mutation rules in DLJSFuzzer. In this table, the first column is the index. The following columns are, respectively, the name, the description, and the type of the mutation rule. According to the different types of tensor operations, the tensor mutation rules in DLJSFuzzer are mainly divided into five categories: tensor copy, tensor padding, tensor transpose, tensor cropping, and data type transformation. Among them, the first four categories are the transformation of the tensor shape, while the last category is the transformation of the tensor data type. More detailed information about these mutation rules is as follows.

\begin{table*}[htpb]
    \centering
    \caption{Tensor Mutation Rules in DLJSFuzzer}
    \begin{tabular}{|c|l|l|l|}
    \hline
       ID  & Mutation Rule & Description & Type\\
    \hline
        1 & Weight Dimension Copy (WDC)& \makecell[l]{Copys the tensor and concatenates along the weight \\dimension to double tensor size} & \multirow{7}{*}{Tensor Copy}\\ \cline{1-3}
        2 & Height Dimension Copy (HDC)& \makecell[l]{Copys the tensor and concatenates along the height \\dimension to double tensor size}&\\ \cline{1-3}
        3 & Channel Dimension Copy (CDC)& \makecell[l]{Copys the tensor and concatenates along the channel\\ dimension to double tensor size}&\\ \cline{1-3}
        4 & Batch Dimension Copy (BDC)& \makecell[l]{Copys the tensor and concatenates along the batch\\ dimension to double tensor size}&\\ \hline
        5 & Weight Dimension Padding (WDP)& Zero pads the tensor along the weight dimension& \multirow{4}{*}{Tensor Padding}\\ \cline{1-3}
        6 & Height Dimension Padding (HDP)& Zero pads the tensor along the height dimension&\\ \cline{1-3}
        7 & Channel Dimension Padding (CDP)& Zero pads the tensor along the channel dimension&\\ \cline{1-3}
        8 & Batch Dimension Padding (BDP)& Zero pads the tensor along the batch dimension&\\ \hline
        9 & \makecell[l]{Height-Weight Dimension \\Transpose (HWDT)}& Transposes the tensor in the height and weight dimension&Tensor Transpose\\ \hline
        10 & Random Cropping (RC)& Randomly crops the tensor to reduce its size& Tensor Cropping\\ \hline
        11 & Float Transformation (FT)& Converts the tensor to the data type $float$& \multirow{3}{*}{\makecell[l]{Data Type \\Transformation}}\\ \cline{1-3}
        12 & Double Transformation (DT)& Converts the tensor to the data type $double$&\\ \cline{1-3}
        13 & BFloat Transformation (BFT)& Converts the tensor to the data type $bfloat16$&\\ \hline
    \end{tabular}
    \label{tensor mutation rule}
\end{table*}

\textbf{Tensor Copy.} The \textbf{Mutation Rule 1-4} belong to the Tensor Copy category. The aim of these mutation rules is to double the size of the tensor along the corresponding dimension. These mutation rules copy the tensor and concatenate the original tensor with the newly copied tensor along the corresponding dimension.

\textbf{Tensor Padding.} The \textbf{Mutation Rule 5-8} belong to the Tensor Padding category. These mutation rules zero pad the tensor along the corresponding dimension. Unlike tensor copy, tensor padding results in a smaller increase in tensor size. The mutation rules of DLJSFuzzer support different granularities of tensor modifications, making the changes to tensor shapes more flexible.

\textbf{Tensor Transpose.} The \textbf{Mutation Rule 9} belongs to the Tensor Transpose category. This mutation rule transposes the tensor along the height and weight dimensions, which is a mutation rule that only changes the shape of the tensor but not the size of the tensor.

\textbf{Tensor Cropping.} The \textbf{Mutation Rule 10} belongs to the Tensor Cropping category. The aim of this mutation rule is to reduce the size of the tensor. The cropped tensor is a part of the original tensor. To ensure the flexibility of DLJSFuzzer in changing tensor shapes, the shape of the cropped tensor can be randomly specified.

\textbf{Data Type Transformation.} The \textbf{Mutation Rule 11-13} belong to the Data Type Transformation category. The aim of these mutation rules is to change the data type of the tensor. Since different data types require different storage space, changing the data type means changing the tensor size. DLJSFuzzer supports three common data types: 32-bit $float$, 64-bit $double$, and 16-bit $bfloat16$.

\subsection{Test Input Model Generation}
\label{test input model generation}
The test input model is another important part of the test input in DL framework testing.
The existing DL framework testing methods lack targeted designs for testing JavaScript DL frameworks. The generated models contain almost no model structures suitable for inference acceleration, making it difficult to effectively detect bugs that may be caused by inference acceleration mechanisms. To overcome this limitation, DLJSFuzzer designs a series of model mutation rules targeting widely used inference acceleration strategies in JavaScript DL frameworks (e.g., Node Optimization, Operator Reordering, and Operator Fusion). DLJSFuzzer first generates a seed model (see Section \ref{seed model generation}), and then applies a mutation rule to generate a new model based on the seed model (see Section \ref{mutation}).

\subsubsection{Seed Model Generation}
\label{seed model generation}
Existing DL framework testing methods usually (e.g. Muffin\cite{muffin}, LEMON\cite{lemon}) use publicly available models as seed models, and generate new models by making small modifications to these seed models through a small number of mutations. The small number of mutations restricts the exploration space of model structures, resulting in generated models that are too similar to the seed models. In order to generate more diversified models, DLJSFuzzer directly generates relatively simple models as the seed model and applies a higher number of mutations to the same seed model. Next, we will introduce the model representation in DLJSFuzzer, and then use this model representation to describe the structure of the seed model.

\textbf{Model Representation.} DLJSFuzzer uses a Directed Acyclic Graph (DAG) to represent each model. The DAG consists of a vertex set $V_G$ and an edge set $E_G$, where each vertex in $V_G$ represents a tensor, and each edge in $E_G$ represents an operator. Various types of edges in the DAG denote different types of operators. When multiple edges converge at the same vertex, the tensor represented by each edge's starting vertex is processed by the corresponding operator, and the output tensors are fused into the tensor represented by the shared endpoint vertex. DLJSFuzzer employs the operator $BatchNorm$ for tensor fusion. In order to maintain the integrity of the test input models, the DAG should adhere to two constraints. Firstly, as test input models should have only one input tensor and one output tensor, the DAG of the models should contain only one source vertex and one sink vertex. Secondly, to ensure the successful construction of generated models, the DAG of the model should be a connected graph.

\textbf{Structure of the Initial Seed Model.} The DAG of the initial seed model in DLJSFuzzer is an operator chain from the source vertex to the sink vertex, which is only composed of the operator $identity$. The length of this operator chain determines the size of the newly generated model, and it is predetermined before the start of the method. In addition, all newly generated models are regarded as seed models in the following rounds.

\subsubsection{Mutation Rules}
\label{mutation}

DLJSFuzzer generates new models by mutating the seed model generated in Section \ref{seed model generation}. To effectively detect quality issues in JavaScript DL frameworks caused by inference acceleration mechanisms, DLJSFuzzer designs eight model mutation rules targeting widely used inference acceleration strategies (e.g., Node Optimization, Operator Reordering, and Operator Fusion), as shown in Table \ref{model mutation rule}. In this table, the first column represents the index, while the following three columns respectively represent the mutation rule, the corresponding description, and the inference optimization strategy the mutation rule targets. Next, we will introduce these mutation rules in detail.

\begin{table*}[htpb]
    \caption{Model Mutation Rules in DLJSFuzzer}
    \centering
    \begin{tabular}{|c|l|l|l|}
    \hline
       ID  & Mutation Rule & Description & \makecell[l]{Targeted Inference \\Acceleration Strategy}\\
    \hline
        1 & Zero-Dimensional Tensorization (ZDT) & Inserts an operator containing zero-dimensional tensor & \multirow{4}{*}{Node Optimization}\\ \cline{1-3}
        2 & In-Place Transpose (IPT) & \makecell[l]{Inserts the operator $Transpose$ without data movement}
&\\ \cline{1-3}
        3 & Height-Weight ReduceMean (HWR)& \makecell[l]{Inserts the operator $ReduceMean$ only on the height\\ and weight dimensions} &\\ \hline
        4 & Transpose-based Operator Reordering (TOR)& \makecell[l]{Inserts the operator combination based on the \\equivalence relation  $A^T \times B^T = {(B \times A)}^T$} & \multirow{3}{*}{Operator Reordering}\\ \cline{1-3}
        5 & Maxpool-based Operator Reordering (MOR)& \makecell[l]{Inserts the operator combination ($MaxPool$, $Sigmoid$) or \\($MaxPool$, $Softmax$)}&\\ \hline
        6 & Equivalent Separable Convolution (ESC)& \makecell[l]{Inserts the operator combination consisting of two depth\\-wise convolutions equivalent to a separable convolution}& \multirow{3}{*}{Operator Fusion}\\ \cline{1-3}
        7 & Equivalent CBR Block (ECB) & \makecell[l]{Inserts the operator combination consisting of three \\operators: $(Convolution, BatchNorm, RELU)$}&\\ \hline
        8 & Random Operator Replace (ROR) & Replaces an operator with a randomly selected operator & None \\ \hline
    \end{tabular}
    \label{model mutation rule}
\end{table*}

\textbf{Mutation Rule 1-3.} These mutation rules (including ZDT, IPT, and HWR) are proposed targeting at Node Optimization. In order to accelerate inference, the JavaScript DL framework replaces high computational resource-consuming nodes with equivalent nodes that have lower computational resource consumption. To effectively trigger these inference acceleration strategies, DLJSFuzzer designs mutation rules to insert high computational resource-consuming nodes in the model. Specifically, in \textbf{Rule 1} (ZDT), DLJSFuzzer inserts an operator containing a zero-dimensional tensor in the model to induce the JavaScript DL framework to replace the zero-dimensional tensor in this operator with a constant. In \textbf{Rule 2} (IPT), DLJSFuzzer inserts an operator $Transpose$ that does not move data to induce the JavaScript DL framework to replace this operator with an operator $Reshape$ that provides the same computation results but executes faster. In \textbf{Rule 3} (HWR), DLJSFuzzer inserts an operator $ReduceMean$ that is only on the height and weight dimensions to induce the JavaScript DL framework to replace this operator with an equivalent operator $AveragePool$.

\textbf{Mutation Rule 4 \& 5.} These mutation rules (including TOR and MOR) are proposed targeting at Operator Reordering. As introduced in Section \ref{background inference acceleration}, many operator combinations satisfy equivalence relations (e.g., commutativity, associativity, and distributivity). To simplify computation and accelerate inference, the JavaScript DL framework rearranges operators in DL model based on these equivalence relations. Mutation 4 \& 5 insert operator combinations that satisfy the above mentioned equivalence relations in the model in the order before reordering, thereby inducing the JavaScript DL framework to rearrange the operator combination. Specifically, \textbf{Rule 4} (TOR) targets the equivalence relation $A^T \times B^T \iff (B \times A)^T$, inserting the operator combination $A^T \times B^T$ into the model. \textbf{Rule 5} (MOR) targets the equivalence relation $(Maxpool, Sigmoid) \iff (Sigmoid Maxpool)$. Since the former executes faster than the latter, DLJSFuzzer inserts the latter operator combination in the model to induce the JavaScript DL framework to reorder it into the former. This Mutation Rule also applies to the equivalence relation $(Maxpool, Softmax) \iff (Softmax, Maxpool)$.

\textbf{Mutation Rule 6 \& 7.} These mutation rules (including ESC and ECB) are proposed targeting Operator Fusion. As introduced in Section \ref{background inference acceleration}, in addition to the equivalence relations between operator combinations mentioned above, many operators and operator combinations also satisfy equivalence relations. Because the inference speed of a single operator is usually faster than that of an equivalent operator combination, the JavaScript DL framework often fuses operator combinations that satisfy equivalence relations into a single operator. To effectively trigger Operator Fusion, DLJSFuzzer inserts operator combinations that satisfy equivalence relations into the model. Specifically, in \textbf{Rule 6} (ESC), since a separable convolution is equivalent to two depth-wise convolutions, DLJSFuzzer inserts two consecutive separable convolutions in the model to induce Operator Fusion. In \textbf{Rule 7} (ECB), since operator combination $(Convolution, BatchNorm, RELU)$ is equivalent to a fused CBR (Conv\_Batch\_Relu)  block, but the former consumes higher computational resources, DLJSFuzzer inserts the operator combination $(Convolution, BatchNorm, RELU)$ into the model to induce Operator Fusion.

\textbf{Mutation Rule 8.} This mutation rule (ROR) aims at generating models with various operators. ROR randomly selects an operator in the model to be mutated and replaces the selected operator with another operator. It is worth noting that, the operator $None$ is also a candidate operator. When the operator $None$ in the model is replaced with another operator, it means inserting an operator into the model; when other operators in the model are replaced with the operator $None$, it means deleting the operator from the model.

\subsection{Bug Detection}
\label{bug detection}
Based on the test inputs generated in Section \ref{test input tensor generation} and \ref{test input model generation}, DLJSFuzzer employs differential testing between TensorFlow \cite{tensorflow}, PyTorch \cite{pytorch}, and the most widely used JavaScript DL framework, TensorFlow.js \cite{tfjs}. To mitigate the impact of browser environment differences, DLJSFuzzer deploys TensorFlow.js in two widely used browsers, including Google Chrome and Microsoft Edge. When the same DL model produces different inference results in the two browsers, DLJSFuzzer will separately re-execute the model's inference process in both browsers until the inference results are equal. If the inference results remain different, the model will be discarded. DLJSFuzzer focuses on two categories of bugs: non-numerical bugs and numerical bugs. Specifically, non-numerical bugs involve crashes, which are detected by analyzing execution logs. Numerical bugs involve Not-A-Number (NaN) bugs and inconsistency bugs. Among them, NaN bugs are detected when the inference result in TensorFlow.js contains NaN while other frameworks' inference results do not. The detection of inconsistency bugs is based on the $inconsistency$ between the inference results. The calculation of $inconsistency$ is as follows: \ding{192} Denote the inference result under each DL framework as $result_f$. The $f$ in $result_f$ denotes the frameworks.
\ding{193} Calculate the difference between all $result_f$ as $result\_diff_{ij}$. The $i$ and $j$ in $result\_diff_{ij}$ denote the frameworks.
\ding{194} Calculate the max scalar value in each $result\_diff_{ij}$ as $inconsistency_{ij}$. The $i$ and $j$ in $inconsistency_{ij}$ denote frameworks.

An inconsistency bug is detected when the $inconsistency$ between TensorFlow and TensorFlow.js, and between PyTorch and TensorFlow.js, both exceed the pre-specified threshold $\epsilon$. To ensure fairness and minimize errors from computational randomness, DLMOSA sets the threshold $\epsilon$ to the maximum among all existing methods (e.g., Gandalf \cite{gandalf}, Muffin \cite{muffin}, etc), which is 0.15.

\subsection{Heuristic Guidance}
\label{heuristic guidance}
To continuously improve the effectiveness of framework testing, DLJSFuzzer designs the heuristic guidance to control the test input model generation in the following rounds based on the bug detection performance of newly generated models. In this section, we will introduce the designed heuristic indicator (Section \ref{heuristic indicator}), and its application in seed model selection (Section \ref{seed model selection}) and model mutation rule selection (Section \ref{mutation rules selection}). 

\subsubsection{Heuristic Indicator}
\label{heuristic indicator}
DLJSFuzzer designs the heuristic indicator $fitness$ to quantitatively measure the bug detection effectiveness. The $fitness$ of each newly generated model is separately calculated and recorded, which is used to reflect the bug detection performance. Depending on the categories of bugs detected, DLJSFuzzer supports different $fitness$ calculations. Specifically, when detecting crashes and NaN bugs, $fitness$ is $m_d$. $m_d$ is 0 when no crashes and NaN bugs are triggered, and $m_d$ is the mean value of the test input tensor when these bugs are triggered. When detecting inconsistency bugs, $fitness$ is $inconsistency$ (see Section \ref{bug detection}).

\subsubsection{Heuristic Guidance in Seed Model Selection}
\label{seed model selection}

To further enhance the effectiveness of the DL framework testing, DLJSFuzzer uses a tournament algorithm \cite{tournament} to select the model with the best bug detection performance from the initial seed model and all models generated in previous rounds as the seed model for the next round of mutation (the first round directly selects the initial seed model). The bug detection performance of the model is measured by the heuristic indicator $fitness$ (introduced in Section \ref{heuristic indicator}).

\subsubsection{Heuristic Guidance in Model Mutation Rules Selection}
\label{mutation rules selection}
Heuristically selecting mutation rules helps to continuously enhance the effectiveness of DLJSFuzzer in detecting bugs in DL frameworks. DLJSFuzzer applies the heuristic indicator $fitness$ to control the probability of each mutation rule being selected.  Based on $fitness$, the probability $p$ of each mutation rule being selected is calculated as follows. Firstly, DLJSFuzzer calculates the total contribution $c$ of each mutation rule:
\begin{center}
    $c_1 = c_0 + \Delta fitness$
\end{center}
where $c_1$ and $c_0$ is the total contribution $c$ after and before mutation, respectively. $\Delta fitness = fitness_{new} - fitness_{old}$, where $fitness_{new}$ is the $fitness$ of the newly generated model and $fitness_{old}$ is the $fitness$ of the model before mutation. Based on $c$, the probability $p$ of each mutation rule being selected is calculated as follows:
    $$p = \frac{c}{\sum_{i=1}^{n}c_i}$$
where $n$ is the number of mutation rules. It is worth noting that, in order to generate models with various operators, when calculating $n$, DLJSFuzzer treats the replacement of each operator with \textbf{Rule 8} (ROR) (see Section \ref{test input model generation}) as a mutation rule.
\section{Evaluation}
To evaluate the bug detection effectiveness of DLJSFuzzer, we conduct an experiment to compare DLJSFuzzer with three state-of-the-art methods, including LEMON\cite{lemon}, Muffin\cite{muffin}, and Gandalf\cite{gandalf}. All the code and experimental results have been open-sourced in our Github repository.

\subsection{Experimental Setup}
\label{experimental Setup}

Existing DL framework testing methods do not support detecting JavaScript DL frameworks (e.g., TensorFlow.js \cite{tfjs}). For the fairness of our experiment, all methods in our evaluation are configured to use the same parameter and code to call the API of the JavaScript DL framework. We separately download the source code of all baselines and run them to generate test input tensors and test input models. After that, we execute the test inputs generated by each method in the same experimental environment. Each method is executed for three hours respectively. The initial seeds are adopted as specified in each method's paper. The workstation's operating system is Ubuntu 22.04, equipped with an Intel Core Processor (Skylake) (16 cores, 2.0GHz). The framework being detected is TensorFlow.js, and other frameworks assisting in the differential testing include TensorFlow \cite{tensorflow} and PyTorch \cite{pytorch}. The versions of each framework are: TensorFlow.js 3.19.0, TensorFlow 2.9.0, PyTorch 1.12.0. TensorFlow.js is deployed on two widely-used browsers, Microsoft Edge and Google Chrome. The browser versions are: Microsoft Edge 125.0.2535.67 (64 bit) and Google Chrome 125.0.6422.112 (64 bit). The threshold value $\epsilon$ for detecting NaN \& Inconsistency bugs is set to 0.15 (see Section \ref{bug detection}). DLJSFuzzer not only supports randomly generating test input tensors but also loading test input tensors from open-source datasets. DLJSFuzzer supports all datasets used by baselines, including: MNIST, Fashion-MNIST, CIFAR-10, ImageNet, Sine-Wave, and Stock-Price.

\subsection{Research Questions}
In our evaluation, we conduct the experiment to answer the following four research questions.

\begin{itemize}
    \item \textbf{RQ1:} How does DLJSFuzzer perform in detecting JavaScript DL framework bugs?
    \item \textbf{RQ2:} What is the root cause of detected JavaScript DL framework bugs?
    \item \textbf{RQ3:} How does DLJSFuzzer perform in efficiency and the diversity of generated test input models?
    \item \textbf{RQ4:} To what extent do the proposed test input tensor and model generation method contribute to bug detection? 
\end{itemize}

In RQ1, we evaluate the effectiveness of DLJSFuzzer when detecting Crashes and NaN \& Inconsistency bugs. In RQ2, we analyze the root cause of all detected crashes. In RQ3, we evaluate the method efficiency and the diversity of test input models (representing the test sufficiency). In RQ4, we conduct an ablation study to investigate the contribution of the proposed test input tensor mutation and test input model mutation.

\subsection{RQ1: Effectiveness in Detecting Bugs}
\label{RQ1_effectiveness}
DLJSFuzzer detects both numerical bugs (including crashes) and non-numerical bugs (including NaN \& Inconsistency bugs) in JavaScript DL frameworks. As mentioned in Section \ref{bug detection}, crashes are confirmed by analyzing the model's runtime logs. The NaN \& Inconsistency bugs are confirmed based on the inference results under different frameworks. Specifically, if there is NaN in the inference result under TensorFlow.js, but no NaN under TensorFlow or PyTorch, an NAN bug is successfully detected. If the max difference between the inference results exceeds the specified threshold $\epsilon$, an inconsistency bug is successfully detected. To avoid redundancy, DLJSFuzzer implements the following post-processing measures: For crash detection, two researchers independently analyze the model's runtime logs to eliminate identical crashes. For NaN \& Inconsistency bug detection, DLJSFuzzer records the model structure that triggers the bug. If the model structures that trigger the bug are identical, the two bugs are considered the same. The number of unique bugs detected by DLJSFuzzer in TensorFlow.js in all datasets is shown in Table \ref{effectiveness_table_effectiveness}. The first column in the table represents the datasets used. The last two columns represent the number of crashes detected and the number of NaN \& Inconsistency bugs detected, respectively. The last row in the table represents the total number of bugs detected by DLJSFuzzer in all datasets after removing redundancy. This table shows that DLJSFuzzer successfully detects 21 unique crashes and 126 unique NaN \& Inconsistency bugs. All crashes have been reported to the open-source community, with 12 of them already confirmed by developers, and the rest are still being processed.

\begin{table}[htpb]
    \caption{Number of Unique Detected Bugs in Each Dataset}
    \centering
    \begin{tabular}{|c|c|c|}
    \hline
         Dataset & Crash & NaN \& Inconsistency\\ \hline
         Random & 18 & 69\\ \hline
         MNIST & 11 & 1\\ \hline
         Fashion-MNIST & 10 & 13\\ \hline
         CIFAR-10 & 12 & 0\\ \hline
         ImageNet & 19 & 88\\ \hline
         Sine-Wave & 8 & 2\\ \hline
         Stock-Price & 8 & 3\\ \hline
         Total & 21 & 126\\ \hline
    \end{tabular}
    \label{effectiveness_table_effectiveness}
\end{table}

To further evaluate the effectiveness of DLJSFuzzer, we compare the number of unique bugs detected by DLJSFuzzer and all baselines (including LEMON, Muffin, and Gandalf). The experimental results are shown in Table \ref{effectiveness_table_baseline}. The first column represents the method name, and the following two columns represent the number of detected crashes and NaN \& Inconsistency bugs, respectively. The experimental result shows that DLJSFuzzer successfully detects 21 crashes and 126 NaN \& Inconsistency bugs, which is the most compared to all baselines. LEMON only detects four NaN \& Inconsistency bugs and does not detect any crashes, which is limited by the low diversity of generated models. Muffin detects 12 crashes and 2 NaN \& Inconsistency bugs. The latter number is limited due to the lack of effective test input tensor mutation. The models generated by Gandalf are too simple, thus failing to trigger any bugs successfully. In addition, we analyze the containment relationship and the root causes of the bugs detected by these methods (see Section \ref{RQ2_root cause} for more introduction about these root causes). We find that the root cause of all crashes detected by Muffin is cache reuse, which is completely covered by our method. The root cause of all NaN \& Inconsistency bugs detected by LEMON and Muffin is random and precision, also completely covered by our method.

\begin{table}[htpb]
    \caption{Comparison in the Number of Unique Detected Bugs}
    \centering
    \begin{tabular}{|c|c|c|}
    \hline
         Method & Crash & NaN \& Inconsistency\\ \hline
         DLJSFuzzer & 21 & 126\\ \hline
         LEMON & 0 & 4\\ \hline
         Muffin & 11 & 2\\ \hline
         Gandalf & 0 & 0\\ \hline
    \end{tabular}
    \label{effectiveness_table_baseline}
\end{table}

\begin{center}
\fcolorbox{black}{lightgray}{\parbox{.95\linewidth}{\textit{Answer to RQ1:} DLJSFuzzer successfully detects 21 unique crashes and 126 unique NaN \& inconsistency bugs in TensorFlow.js, which is the most among existing baselines and completely covers all bugs detected by existing baselines. The experimental results fully demonstrate the superior effectiveness of DLJSFuzzer in detecting bugs in JavaScript DL frameworks.}}
\end{center}

\subsection{RQ2: Root Cause}
\label{RQ2_root cause}
As shown in Table \ref{effectiveness_table_effectiveness}, DLJSFuzzer successfully detects 21 unique crashes and 126 NaN \& Inconsistency bugs. In this section, we will further analyze the root cause of these bugs as follows.

\subsubsection{Root Cause of Crashes.}
\begin{figure}[htpb]
    \centering
    \includegraphics[scale=0.18]{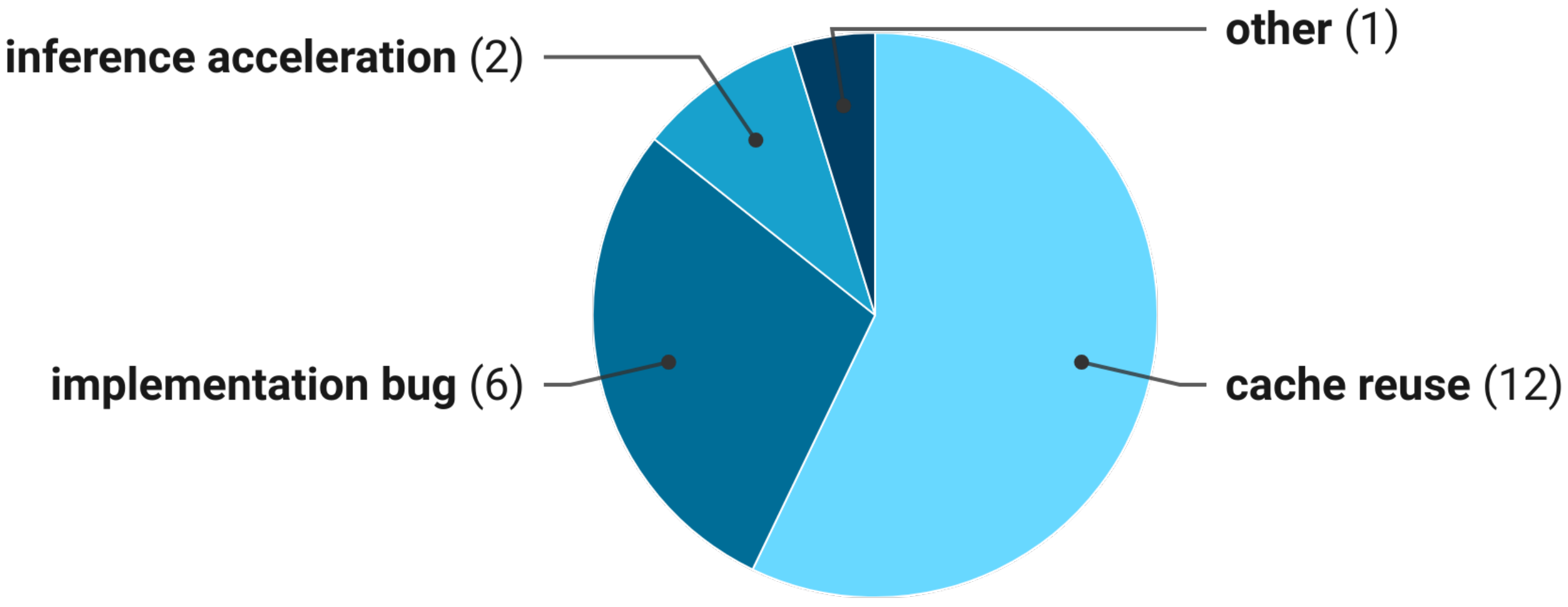}
    \caption{Root Cause of Detected Crashes}
    \label{evaluation_root_cause_crash}
\end{figure}

By analyzing the execution logs of detected crashes, DLJSFuzzer categorizes the root causes of detected crashes into four types: cache reuse, implementation bug, inference acceleration, and others. The number of crashes in each category is illustrated in Figure \ref{evaluation_root_cause_crash}.

\textbf{Cache reuse (12/21, 57.14\%).} Cache reuse is the most common root cause found in the detected crashes. This type of bug primarily stems from inconsistencies in tensor dimensions. More specifically, flaws in the cache reuse mechanism result in the loss of certain parameters within the tensor, consequently triggering alterations in the tensor dimensions. The bug occurs when the actual tensor dimensions received by the operator do not align with the declared dimensions.

\textbf{Implementation bug (6/21, 28.57\%).} This type of bug primarily stems from certain operators in TensorFlow.js lacking support for specific data types or parameter settings. For instance, the operator $Batch\_normalization$ in TensorFlow.js does not support the data type $bfloat16$. The bug occurs when a tensor of data type $bfloat16$ is passed to this operator.

\textbf{Inference acceleration (2/21, 9.52\%).} This type of bug primarily stems from parameter setting errors in TensorFlow.js during inference acceleration. Specifically, TensorFlow.js incorrectly sets parameters for optimized operators (or operator combinations) while applying strategies such as node optimization and operator fusion, leading to framework bugs.

\textbf{Other (1/21, 4.76\%).} There are still some other root causes of detected bugs. For example, due to the weakly-typed nature of JavaScript, occasional data type conflicts may arise when passing parameters within TensorFlow.js, leading to framework bugs.

\subsubsection{Root Cause of NaN \& Inconsistency Bugs.}

\begin{figure}[htpb]
    \centering
    \includegraphics[scale=0.13]{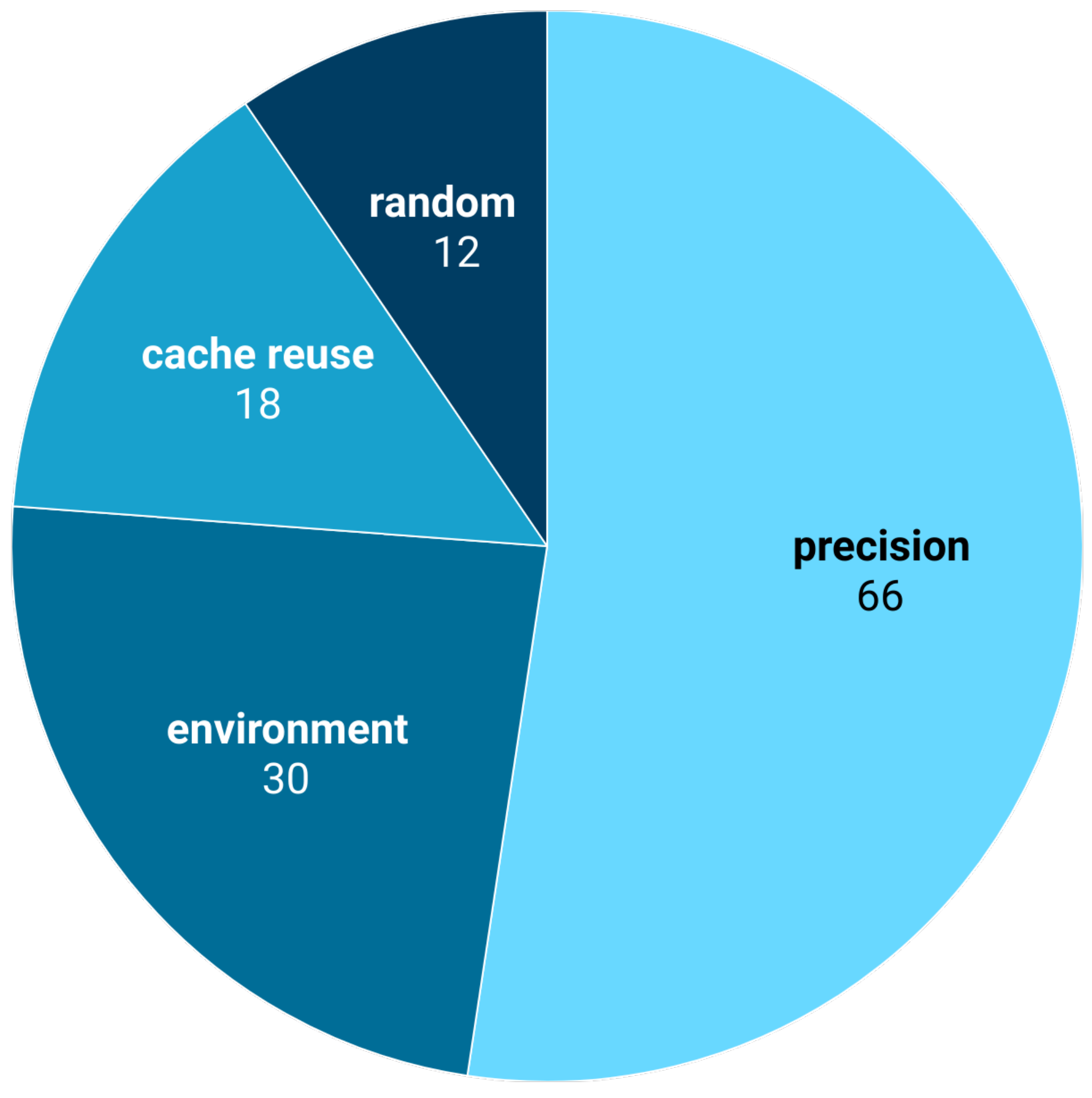}
    \caption{Root Cause of Detected NaN \& Inconsistency Bugs}
    \label{evaluation_root_cause_inconsistency}
\end{figure}

By reproducing and locating detected NaN \& Inconsistency bugs, DLJSFuzzer categorizes root causes of detected NaN \& Inconsistency bugs into four types: precision, environment, cache reuse, and random. Figure \ref{evaluation_root_cause_inconsistency} shows the number of NaN \& Inconsistency bugs in each category.

\textbf{Precision (66/126, 52.38\%).} Precision loss is the most common root cause of NaN \& Inconsistency bugs detected by DLJSFuzzer. On the one hand, large-scale numerical calculation inevitably comes with precision loss. On the other hand, the weakly-typed nature of JavaScript exacerbates precision loss. Precision loss accumulates during model inference, leading to framework bugs.

\textbf{Environment (30/126, 23.81\%).} This type of bug primarily stems from the fluctuation of the framework deployment environment (especially browser environment). Compared to the local environment, the browser environment in which TensorFlow.js is deployed is more unstable and susceptible to factors such as network and operational context, leading to framework bugs.

\textbf{Cache Reuse (18/126, 14.29\%).} The flawed cache reuse mechanism leads to framework bugs. Specifically, the flawed cache reuse occasionally causes parameters to be incorrectly overwritten or modified, resulting in framework bugs.

\textbf{Random (12/126, 9.52\%).} The introduction of randomness by some operators under test also leads to some framework bugs, for example, the operator $Dropout$.

\begin{center}
\fcolorbox{black}{lightgray}{\parbox{.95\linewidth}{\textit{Answer to RQ2:} In DLJSFuzzer, the root causes of detected crashes include cache reuse, implementation bug, inference acceleration, and other. The root causes of detected NaN \& Inconsistency bugs include precision, environment, cache reuse, and random. }}
\end{center}

\subsection{RQ3: Efficiency \& Model Diversity}
\label{RQ3_efficiency}

In this section, we evaluate the efficiency of DLJSFuzzer and the diversity of generated test input models. Model diversity is a widely used measurement to reflect the sufficiency of framework testing. The higher the model diversity, the more comprehensive the framework testing. Model diversity is calculated by the indicator $med$ (mean edit distance). $med$ is defined as the average of the edit distances between all pairs of models. The edit distance between two models is the minimum value of operations needed to transform one model into another. The comparison of efficiency is shown in Table \ref{effectiveness_figure_efficiency}. The first column represents the method name, and the second column represents the average time consumption for each round of framework testing. The last column represents the average time consumption for detecting each framework bug. Due to Gandalf not detecting any bugs, it is unable to calculate the average time Gandalf takes to detect each bug, so we mark it as NaN in the table. From this table, we can conclude that compared to all baselines, DLJSFuzzer has the shortest testing time per round (19.60s). Compared to the best performance in the baselines (36.99s per round), the model generation efficiency improvement is over 47\%. Besides, DLJSFuzzer has the shortest time to detect each bug (73.47s). Compared to the best performance in the baselines (830.77s per bug), the bug detection efficiency improvement is over 91\%. The improvement in the efficiency of DLJSFuzzer is attributed to the models generated by DLJSFuzzer effectively triggering the inference acceleration mechanism.

\begin{table}[htpb]
    \centering
    \caption{Comparison in Efficiency}
    \begin{tabular}{|c|c|c|}
    \hline
      Method  & Time Per Round (s) &  Time Per Bug (s)\\ \hline
      DLJSFuzzer & 19.60 & 73.47\\ \hline
      LEMON  & 63.91 &  2,700\\ \hline
      Muffin & 47.79 &  830.77\\ \hline
      Gandalf & 36.99 & NaN\\ \hline
    \end{tabular}
    \label{effectiveness_figure_efficiency}
\end{table}

\begin{figure}[htpb]
    \centering
    \includegraphics[scale=0.13]{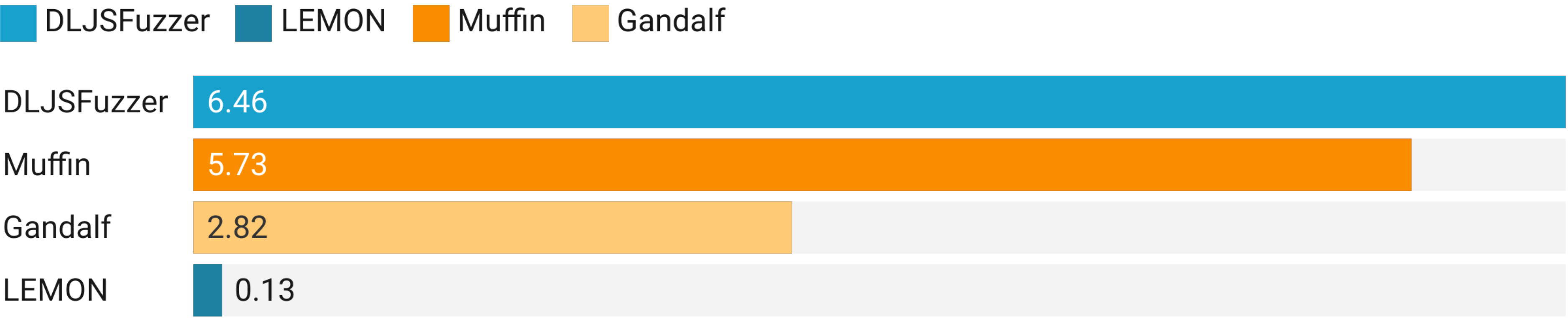}
    \caption{Comparison in Model Diversity}
    \label{effectiveness_figure_diversity}
\end{figure}

To calculate the model diversity, we save all models generated by each method in three hours and compute their $med$. The experimental results are shown in Figure \ref{effectiveness_figure_diversity}. This figure uses different colors to represent different methods, with the numbers in the figure indicating the $med$ of the models generated by each method. The figure shows that the $med$ of the models generated by DLJSFuzzer is 6.46, higher than 5.73 in Muffin, 2.82 in Gandalf, and 0.13 in LEMON. We can conclude that the model mutation in DLJSFuzzer also contributes to increasing model diversity.

\begin{center}
\fcolorbox{black}{lightgray}{\parbox{.95\linewidth}{\textit{Answer to RQ3:} Compared to three baselines, DLJSFuzzer has grown by over 47\% in model generation efficiency and over 91\% in bug detection efficiency. In addition, the models generated by DLJSFuzzer have the highest model diversity, reflecting the framework testing in DLJSFuzzer is more sufficient.}}
\end{center}

\subsection{RQ4: Ablation Study}
\label{ablation study}
The test input model mutation and test input tensor mutation are two key components of DLJSFuzzer. In this section, we conduct an ablation study to explore the contribution of these two components to bug detection in JavaScript DL frameworks. Specifically, we set up three baselines named $DLJSFuzzer_m$, $DLJSFuzzer_t$, and $DLJSFuzzer_n$. In $DLJSFuzzer_m$, test input tensor mutation is disabled, and all test input tensors are generated randomly. In $DLJSFuzzer_t$, test input model mutation is disabled, and all models are initial seed models (introduced in Section \ref{seed model generation}). In $DLJSFuzzer_n$, both test input tensor mutation and test input model mutation are disabled. The number of bugs detected by DLJSFuzzer and all baselines are shown in Table \ref{effectiveness_table_ablation_study}. The meanings of each column in Table \ref{effectiveness_table_ablation_study} are the same as those in Table \ref{effectiveness_table_baseline}. From the table, it can be seen that compared to other baselines, DLJSFuzzer detects the most crashes (21) and NaN \& Inconsistency bugs (126). Compared to $DLJSFuzzer_n$, $DLJSFuzzer_m$ detects more crashes (12) and NaN \& inconsistency bugs (44), demonstrating the effectiveness of test input model mutation. In particular, due to the overly simple model structure, $DLJSFuzzer_t$ and $DLJSFuzzer_n$ do not successfully trigger any framework bugs. However, when test input tensor mutation and test input model mutation act together, the test input tensor mutation will exert its effect. Specifically, DLJSFuzzer outperforms $DLJSFuzzer_m$, indicating that test input tensor mutation can further improve the bug detection effectiveness of the method based on test input model mutation.

\begin{table}[htpb]
    \caption{Number of Unique Detected Bugs in Ablation Study}
    \centering
    \begin{tabular}{|c|c|c|}
    \hline
         Method & Crash & NaN \& Inconsistency\\ \hline
         $DLJSFuzzer$ & 21 & 126\\ \hline
         $DLJSFuzzer_m$ & 12 & 44\\ \hline
         $DLJSFuzzer_t$ & 0 & 0\\ \hline
         $DLJSFuzzer_n$ & 0 & 0\\ \hline
    \end{tabular}
    \label{effectiveness_table_ablation_study}
\end{table}

\begin{center}
\fcolorbox{black}{lightgray}{\parbox{.95\linewidth}{\textit{Answer to RQ4:} In the ablation study, DLJSFuzzer outperforms $DLJSFuzzer_m$, and $DLJSFuzzer_m$ outperforms $DLJSFuzzer_n$. This result shows our test input model mutation is effective, and the test input tensor mutation can further improve the bug detection effectiveness together with test input model mutation.}}
\end{center}
\section{Validity \& Threat}
In this section, we will analyze the threats to validity in this paper, including internal threats, external threats, and construct threats.

The internal threats lie in the experimental configuration. On one hand, to reduce the impact of experimental randomness, DLJSFuzzer generates over 100 models for each dataset respectively. On the other hand, in order to investigate whether the design of tensor generation and model generation in DLJSFuzzer contributes to the improvement of framework testing effectiveness, we conducted an ablation experiment to explore the effects of each component of DLJSFuzzer (see Section \ref{ablation study}).

The external threats lie in the choice of browsers, and the tested JavaScript DL framework and optimization strategies. Firstly, DLJSFuzzer tests one JavaScript DL framework, TensorFlow.js. Although there are other JavaScript DL frameworks available (e.g., WebDNN \cite{webdnn}, etc.), to our knowledge, TensorFlow.js is currently the only popular and actively maintained JavaScript DL framework. The core idea of DLJSFuzzer is also applicable to testing other JavaScript DL frameworks. Secondly, the choice of browsers may threaten the validity of our method. To mitigate the impact of browser differences on experimental results, we repeatedly executed the model under two widely used browsers, Microsoft Edge and Google Chrome, until the inference results were consistent. Thirdly, tested optimization strategies may threaten the validity of our method. DLJSFuzzer can also be generalized to other optimizations (e.g., Dead Code Elimination) by applying more tensor and model mutation rules.

The construct threats lie in the root cause analysis and the selection of baselines. In the root cause analysis, DLJSFuzzer identifies crashes and analyzes their root causes by manually reviewing execution logs. To minimize the impact of subjective factors on the experimental results, two researchers independently review the execution logs. If the analyses of the root cause of a crash by the two researchers are inconsistent, a third expert will determine the final root cause. In addition, DLJSFuzzer further analyzes the root cause of all confirmed bugs (see our Github for more details). In the selection of baselines, as a model-based DL framework testing method, DLJSFuzzer adopts state-of-the-art model-based baselines (e.g., Muffin \cite{muffin}, Gandalf \cite{gandalf}, etc.), and does not adopt interface-based baselines (e.g., FreeFuzz \cite{freefuzz}, DeepREL \cite{DeepREL}, etc.). In fact, existing interface-based baselines do not consider the unique optimization mechanisms in JavaScript DL frameworks, and therefore cannot effectively detect JavaScript DL framework bugs.
\section{Related Works}

\subsection{Testing DL Frameworks \& Compilers}

DL frameworks and compilers serve as the underlying support for DL applications. Their quality has attracted widespread attention from researchers. Testing methods for various DL frameworks and compilers have been proposed. Some methods aim to detect quality issues in DL compilers such as TVM \cite{tvm}, with representative works like Nnsmith \cite{nnsmith}, TVMfuzz \cite{TVMfuzz}, MTDLComp \cite{mtdlcomp}, Neuri \cite{NeuRI}. They only target conventional optimization mechanism in general DL compilers (e.g., TVM) and do not include JavaScipt-specific inference acceleration strategies or their unique implementations. Therefore, they cannot effectively detect bugs in TensorFlow.js. Some other methods focus on detecting quality issues in common DL frameworks such as TensorFlow and PyTorch. Among them, Audee \cite{audee} is one of the early representative works that introduce differential testing into DL framework testing. Predoo \cite{predoo} is the first to test precision bugs in DL frameworks (e.g., NaN \& Inconsistency). FreeFuzz \cite{freefuzz} improves the test oracle of differential testing. DeepREL \cite{DeepREL} is the first to reveal the equivalence relationship between operators and operator combinations in DL models. Cradle \cite{cradle} proposes using DL models as test inputs. Graph-Fuzz \cite{fanggraphbased} designs the overall workflow of a generation-based model generation method. LEMON \cite{lemon} designs the overall workflow of a mutation-based model generation method. Based on Graph-Fuzz and LEMON, researchers have designed various model generation methods to improve the diversity of test input models. Among them, Muffin \cite{muffin} further refines the generation-based DL framework testing, expanding the model search space from simple model structures to model parameters. Gandalf \cite{gandalf} inherits and improves the mutation-based model generation method, introducing various RL techniques (e.g., Q-learning \cite{qlearning}, DQN \cite{dqn}, DDQN \cite{ddqn}) to make the generated models more diversified. Ramos \cite{Ramos} designs a hierarchical heuristic model mutation method. COMET \cite{comet} introduces heuristic guidance in mutation-based methods, improving the diversity of generated models. However, the above methods fail to consider the resource scarcity in JavaScript environment and corresponding optimization mechanisms, so it is not suitable for detecting JavaScript DL frameworks. Our approach follows the overall workflow of mutation-based DL framework testing, and designs tensor mutation rules and model mutation rules targeting the optimization mechanisms in JavaScript DL frameworks. In conclusion, our method expands the applicability of mutation-based DL framework testing methods to the JavaScript environment.

\subsection{Bugs in DL Frameworks}
As a significant threat to the quality of DL applications, bugs in DL frameworks have received extensive attention from researchers in recent years.
Zhang \cite{DLbug1} conducts an empirical study on user coding errors in TensorFlow-based applications. Based on Zhang's work, M.J. \cite{DLbug2,DLbug3} studies five popular DL frameworks focusing on bugs such as incorrect parameters, invalid structures, and API abuse. Hummatova \cite{DLbug4} focuses on the DL framework bugs related to the training process. Zhong \cite{DLbug5,DLbug6} studies the internal bugs of TensorFlow. Du \cite{DLbug7} explores triggering conditions to classify DL framework bugs, and then analyzes frequency distribution and evolution characteristics of different bug types. Furthermore, Feng \cite{DLbug8} summarizes the classification, symptoms, root causes, fix methods, and fix costs of DL framework bugs at the source code level, then divides DL framework bugs into 11 categories, symptoms into five categories, and root causes into 13 categories.

In recent years, JavaScript DL applications in web browsers have been increasingly popular \cite{DLJSapplication1,DLJSapplication2,DLJSapplication3}. Despite there are few studies on bugs in JavaScript DL frameworks, the quality issues of JavaScript DL frameworks have attracted widespread attention from researchers. Among them, Guo \cite{DLJSbug3} conducts an empirical study to explore the prediction accuracy and performance of trained models when they are transferred from PCs to web browsers. Ma \cite{DLJSbug2} identifies the performance gaps between DL in browsers and on native platforms by comparing the performance of TensorFlow.js and TensorFlow in Python. Quan \cite{DLJSbug1} collects 700 real-world faults from JavaScript-based DL systems and analyzes their fault symptoms, root causes, and fix patterns, respectively. The above study summarizes the characteristics and root causes of bugs in JavaScript DL frameworks, providing inspiration for our design of JavaScript DL framework testing methods.

\section{Conclusion}
\label{conclusion}

In this work, we propose DLJSFuzzer, which is the first to target JavaScript DL optimization mechanisms in framework testing. To comprehensively test components associated with cache reuse in JavaScript DL frameworks, DLJSFuzzer designs 13 tensor mutation rules to generate new test input tensors with flexible shapes and data types. Additionally, to generate new models that effectively trigger the inference acceleration mechanism, DLJSFuzzer designs eight model mutation rules targeting three inference acceleration strategies, including Node Optimization, Operator Reordering, and Operator Fusion. We evaluate DLJSFuzzer to test the most widely used JavaScript DL framework, TensorFlow.js. Experimental results show that DLJSFuzzer successfully detects 21 unique crashes and 126 unique NaN \& Inconsistency bugs. Furthermore, DLJSFuzzer's model generation and bug detection efficiency have increased by more than 47\% and 91\%, respectively. In the future, we will design testing methods for DL frameworks targeting more optimization mechanisms in the JavaScript environment.

\bibliographystyle{acm}
\bibliography{Reference}
\end{document}